%% file: main.tex
\documentclass[a4paper,11pt]{article}
\pdfoutput=1
\usepackage{pos}

\usepackage{graphicx} 
\usepackage{lineno}

\title{\boldmath Electronics design and testing of the CMS Fast Beam Condition Monitor for HL-LHC}

\author*[m]{K. Shibin}
\author[a]{G. Auzinger}
\author[h]{H. Bakhshiansohi}
\author[a]{A. Dabrowski}
\author[i]{A. Dierlamm}
\author[f]{M. Dragicevic}
\author[h]{A. Gholami}
\author[g]{G. Gomez}
\author[c]{M. Guthoff}
\author[a]{M. Haranko}
\author[a]{A. Homna}
\author[m]{M. Jenihhin}
\author[a]{J. Kaplon}
\author[k, a]{O. Karacheban}
\author[b]{B. Korcsm\'aros}
\author[l]{A. Lokhovitskiy}
\author[a]{R. Loos}
\author[i]{S. Mallows}
\author[a]{J. Michel}
\author[c]{V. Myronenko}
\author[d]{G. P\'asztor}
\author[e]{J. Schwandt}
\author[h]{M. Sedghi}
\author[j]{A. Shevelev}
\author[e]{G. Steinbrueck}
\author[j]{D. Stickland}
\author[a]{G.J. Wegrzyn}

\affiliation[a]{CERN, European Organization for Nuclear Research, Geneva, Switzerland}
\affiliation[b]{University of Debrecen, Debrecen, Hungary}
\affiliation[c]{Deutsches Elektronen-Synchrotron, Hamburg, Germany}
\affiliation[d]{ELTE E\"otv\"os Loránd University, Budapest, Hungary}
\affiliation[e]{University of Hamburg, Hamburg, Germany}
\affiliation[f]{HEPHY, Wienna, Austria}
\affiliation[g]{Instituto de Física de Cantabria (IFCA), CSIC-Universidad de Cantabria, Santander, Spain}
\affiliation[h]{Isfahan University of Technology, Isfahan, Iran}
\affiliation[i]{Karlsruher Institut fuer Technologie, Karlsruhe, Germany}
\affiliation[j]{Princeton University, Princeton, NJ, USA}
\affiliation[k]{Rutgers, The State University of New Jersey, Piscataway, NJ, USA}
\affiliation[l]{University of Canterbury, Christchurch, New Zealand}
\affiliation[m]{Tallinn University of Technology, Tallinn, Estonia}

\emailAdd{konstantin.shibin@taltech.ee}

\abstract{
The high-luminosity upgrade of the LHC (HL-LHC) brings unprecedented requirements for precision bunch-by-bunch luminosity measurement and beam-induced background monitoring in real time.
A key component of the CMS Beam Radiation Instrumentation and Luminosity detector system is a stand-alone luminometer, the Fast Beam Condition Monitor (FBCM), which is
able to operate independently at all times with a triggerless asynchronous readout. FBCM utilizes a dedicated front-end ASIC to amplify the signals from CO$_2$-cooled silicon-pad sensors with 1\,ns timing resolution.
Front-end (FE) electronics are subject to high-radiation conditions,
thus all components are radiation hardened: sensors, ASICs, transceivers, etc. 
The FBCM ASIC contains 6 channels, each outputting a high-speed binary signal carrying the time-of-arrival and time-over-threshold information.
This signal is sent via a gigabit optical link to the back-end electronics for analysis.
A dedicated test system is designed for the FBCM FE electronics with a modular setup for all testing needs of the project from initial ASIC validation test to system-level testing with the full read-out chain. The paper reports on the design, read-out architecture, and testing program for the FBCM electronics.
}

\FullConference{6th International Conference on Technology and Instrumentation in Particle Physics (TIPP2023)\\
 4 - 8 Sep 2023\\
Cape Town, Western Cape, South Africa\\}

\begin{document}
\maketitle

\input{tex/fbcm}

\input{tex/readout}

\input{tex/testsystem}

\input{tex/testplan}

\input{tex/summary}

\acknowledgments
We acknowledge the support by the following institutes and funding
agencies: CERN; 
the national research projects RVTT3 "CERN Science Consortium of Estonia", CoE TK202 “Foundations of the Universe” and PUT PRG1467 "CRASHLESS" (Estonia);
Helmholtz-Gemeinschaft Deutscher Forschungszentren (HGF) (Germany); National Research, Development and Innovation Office (NKFIH), including contract numbers K 143460 and TKP2021-NKTA-64 (Hungary);
the US CMS operations program, the US National Science Foundation (NSF), and the US Department of Energy (DOE) (USA).

\input{main.bbl}

\end{document}

%% file: tex/fbcm.tex
\section{Fast Beam Condition Monitor}
\label{sec:fbcm}
Fast Beam Condition Monitor (FBCM) is the next dedicated luminometer being constructed by the Beam Radiation, Instrumentation and Luminosity (BRIL) project of the CMS experiment~\cite{IPRD23, Collaboration:2759074}.
The primary role of the FBCM system is an accurate luminosity measurement in real time and the monitoring of beam-induced background enabled by its 1\,ns time resolution. 
To meet the requirements to achieve a final uncertainty on the integrated luminosity of better than 1\% for the best physics exploitation of the CMS data, the FBCM sensors should have a particular area and distance to the beamline, balancing occupancy, double-hit probability, and acceptance as these determine the linearity of the measurement and its statistical power. 
A good compromise between the area and the position of the sensor in the experiment is provided by 1.7$\times$1.7\,mm$^2$ silicon pad sensors installed at a radius of around 14.5\,cm \cite{Collaboration:2759074}.  
The radiation environment in this position is rather harsh, and the detector modules should withstand up to 200\,Mrad of total ionizing dose and particle fluence up to 2.5$\times$10$^{15}$\,cm$^{-2}$\,1\,MeV neutron equivalent. 

At the core of the FBCM luminometer is the FBCM23 ASIC \cite{TWEPP23}, which is implemented in CMOS 65\,nm process and is designed to withstand the radiation at the installation location.
The FBCM23 ASIC comprises 6 channels of the fast front-end amplifier working in transimpedance configuration, booster amplifier, and leading edge discriminator. The complete processing chain provides an overall shaping function equivalent to a CR-RC$^3$ filter.

FBCM23 ASIC has separate digital and analog power rails powered at 1.2\,V, consuming around 28\,mA and 32\,mA respectively at nominal settings. The slow-control communication with FBCM23 ASIC is organized via an I2C bus. In addition, an analog multiplexer output of the ASIC can be used during validation to evaluate internal analog signals. The internal control registers and bus logic are protected against radiation-induced single-event upsets using triple modular redundancy.

%% file: tex/readout.tex
\section{Readout architecture}
\label{sec:readout}

\begin{figure}[b!]
    \includegraphics[width=\linewidth]{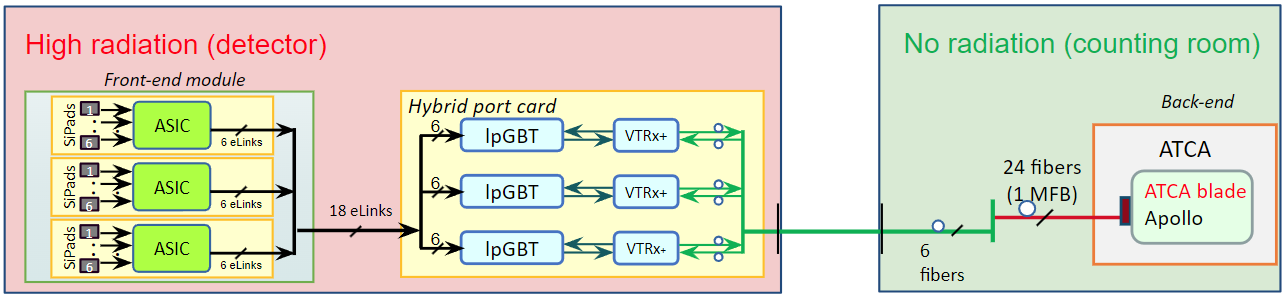}
    \caption{FBCM readout architecture (one service quadrant).}
    \label{fig:readout-architecture}
\end{figure}

The binary architecture employed in the FBCM23 ASIC provides simplification of the off-detector electronics and allows for direct interfacing to the \textit{low-power gigabit transceiver}~\cite{Biereigel:2020jri} (lpGBT) chip, which is the core of the CERN-standard radiation hardened digital transmission system. The general read-out architecture of FBCM is shown in Figure~\ref{fig:readout-architecture} and described in more detail in Ref.~\cite{Collaboration:2759074}.
FBCM is built using four identical half-disks, each containing four identical \emph{service quadrants} which are independent from each other. Each service quadrant contains three identical FE modules, each comprises one FBCM23 ASIC and six sensors.
Each of the ASIC's six channels has an input connected to a CO$_2$-cooled silicon-pad sensor and a differential scalable low-voltage signal output for direct interfacing to the lpGBT transceiver. The input signal is compared to the selectable threshold level and then converted to a binary output signal.
The operation of the FBCM ASIC is asynchronous with respect to LHC clock and triggerless. ASIC output provides a rectangular pulse corresponding to the processed analog signal received from the silicon-pad sensors. The time-of-arrival (ToA) and the time-over-threshold (ToT) parameters are later extracted in the back-end electronics for each particle which generated a signal in the sensors. ToA is the timing relationship between the LHC clock and the sensor signal crossing the user-defined threshold. ToA is aggregated in the histogram for per-bunch-crossing luminosity and background measurement. ToT is used for the sensor stability and performance monitoring over long time.

Because of the limited number of channels on the detector module (6), it is possible to read each single FBCM23 ASIC channel output by the lpGBT input, which is capable of sampling the data with 0.78\,ns time bin resolution. 
Encoded gigabit data from lpGBT is then sent to the back end over an optical link via the \textit{versatile link plus transceivers}~\cite{Soos:2017stv} (VTRx+).
Finally, the data is decoded and analysed in the back-end FPGA-based \textit{Apollo} board developed following the ATCA specifications~\cite{Apollo}. A specialized firmware measures the timing characteristics of the incoming signal and aggregates them into histograms with a bin corresponding to each bunch crossing identifier in the LHC orbit and the data integrated over a time period, called lumi word, of about 1 second.

%% file: tex/testsystem.tex
\section{Test system}
\label{sec:testsystem}

\begin{figure}[b!]
\begin{minipage}[t]{0.45\linewidth}
    \includegraphics[width=\linewidth]{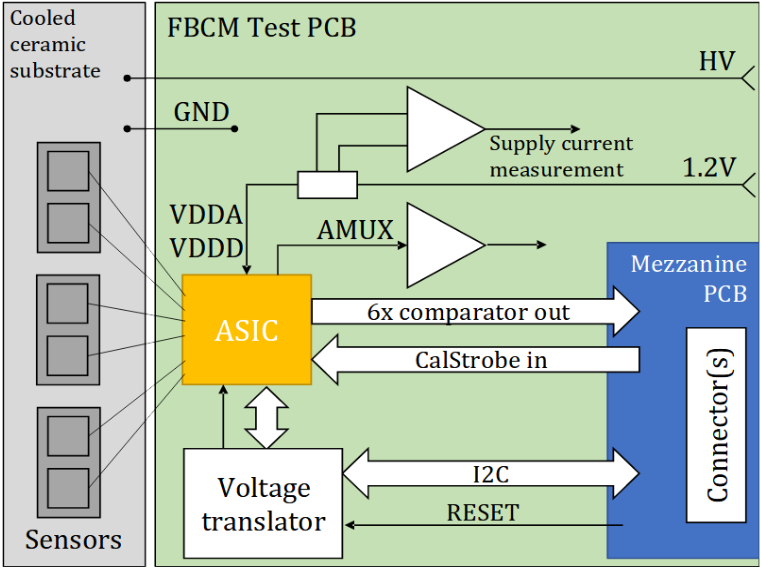}
    \caption{Test system schematic.}
    \label{fig:test-system-block-diagram}
\end{minipage} 
    \hfill%
\begin{minipage}[t]{0.5\linewidth}
    \includegraphics[width=\linewidth]{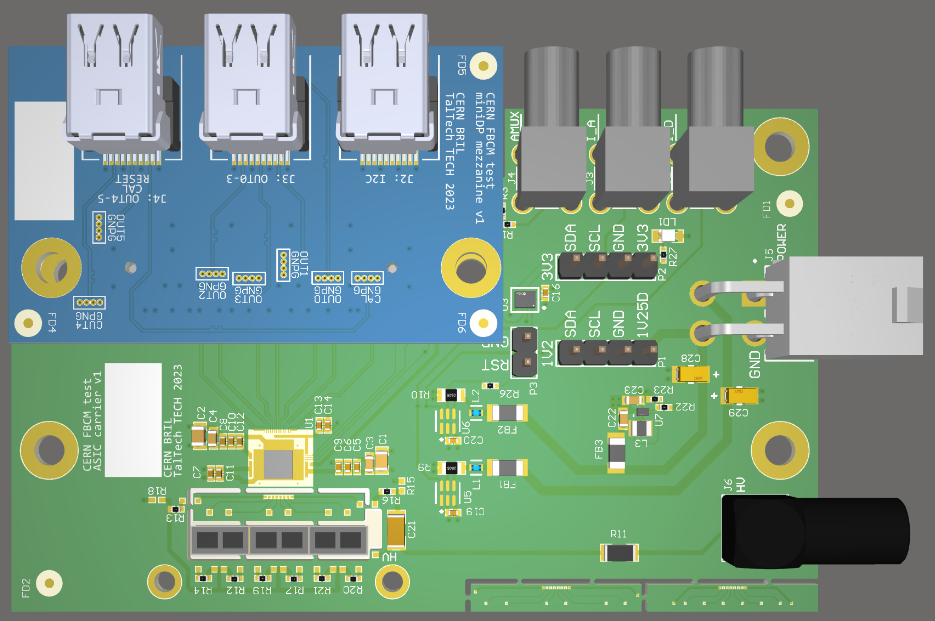}
    \caption{Test system render.}
    \label{fig:test-system-pcb-render}
\end{minipage} 
\end{figure}

As a new design, FBCM23 ASIC requires validation and qualification of its characteristics in different conditions. Initially, basic functionality of the ASIC needs to be evaluated, like I2C communication, digital-to-analog conversion, strobe pulse input and binary signal outputs. At later stages, the ASIC needs to be tested in the laboratory with a radioactive source or in a beam line together with the silicon-pad sensors and the full read-out chain.

For the qualification and validation and all other testing stages of FBCM23 ASIC, we have designed and built a custom test system. Its purpose is to carry an ASIC and sensors, provide convenient connection options to all relevant signals, provide power, and measure current consumption on the ASIC power rails.

The block diagram of the test system is shown in Figure~\ref{fig:test-system-block-diagram}, and a rendering is shown in Figure~\ref{fig:test-system-pcb-render}. It consists of an ASIC carrier board (green PCB), various interface mezzanine boards (blue PCB), and an aluminum nitride (AlN) sensor plate (white, under sensors).

The PCBs are manufactured using standard process, with most off-the-shelf components mounted using automated manufacturing equipment. Afterwards, the ASIC is glued and wire-bonded to the ASIC carrier board. Similarly, silicon-pad sensors are attached to the AlN sensor plate for better cooling contact and then wire-bonded to the pitch adapter area of the ASIC carrier board.

\begin{figure}[b]
\begin{minipage}[t]{0.5\linewidth}
    \includegraphics[width=\linewidth]{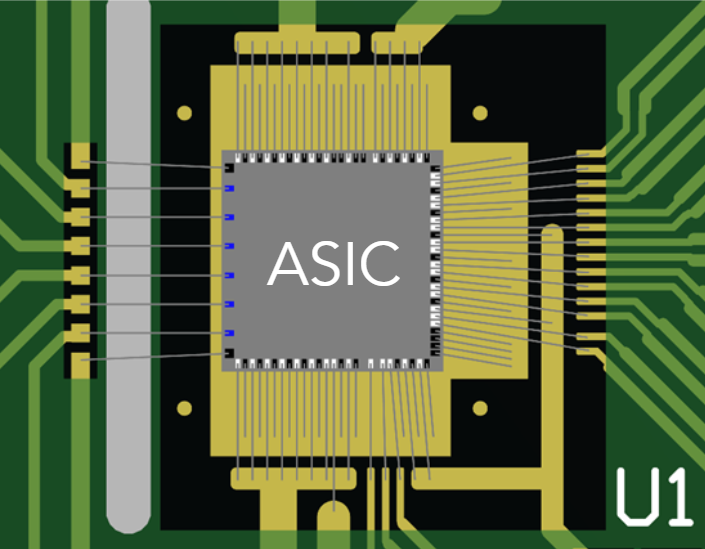}
    \caption{FBCM23 ASIC bonding scheme.}
    \label{fig:asic-bonding}
\end{minipage} 
    \hfill%
\begin{minipage}[t]{0.47\linewidth}
    \includegraphics[width=\linewidth]{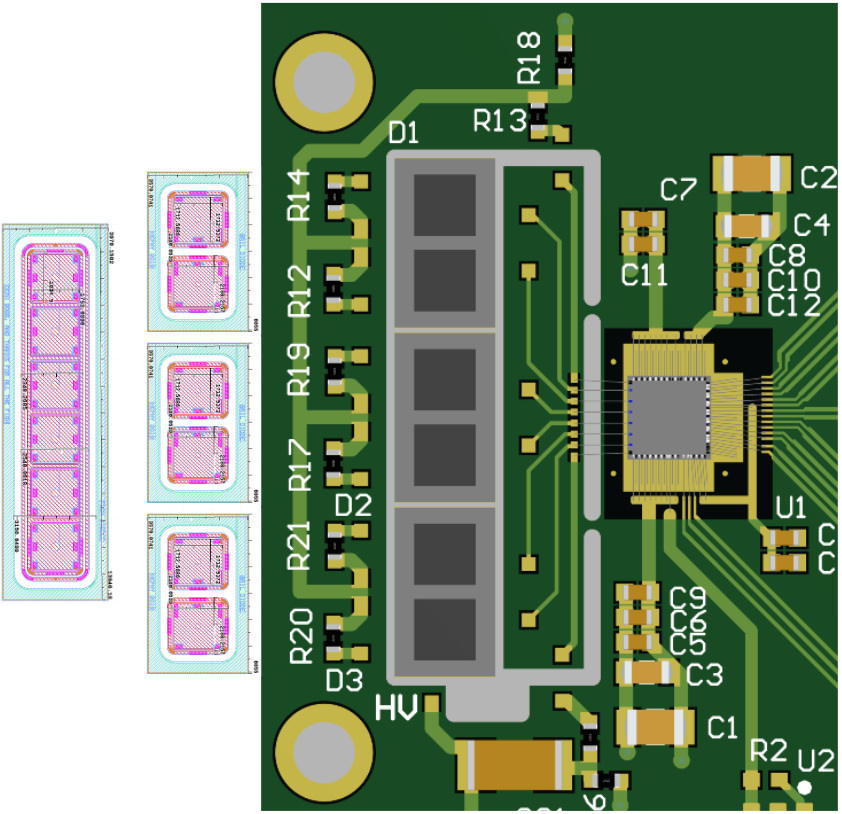}
    \caption{Sensor options matching the FBCM23 ASIC.}
    \label{fig:sensors}
\end{minipage} 
\end{figure}

\subsection{ASIC carrier board}
The ASIC carrier board is a 4-layer 85$\times$53mm PCB. Its main purpose is to host the FBCM23 ASIC and to provide all the necessary connections to it. The main functional parts of the carrier board are as follow.
    \textbf{The ASIC bonding footprint} is the location where the ASIC can be bonded with minimal angle between the bonding wire and the bonding pad longitudinal axis (see Figure~\ref{fig:asic-bonding}).
    \textbf{The ASIC power supply} requires decoupling and is monitored by a current consumption measurement of the digital and analog 1.2V power rails using shunt resistors and specialized amplifiers.
    \textbf{The high voltage supply} includes input via LEMO 00-series connector and an RC-filter.
    \textbf{The pitch adapter is next to the sensor cut-out} which can host two types of silicon-pad sensors (see Figure~\ref{fig:sensors}).
    \textbf{The interface mezzanine connector} (see Section~\ref{sec:mezzanines}) provides flexibility for use in various environments (see Section~\ref{sec:mezzanines}).  
    \textbf{The I2C bus voltage translator} enables communication between ASIC I2C slave interface at 1.2V supply voltage and external I2C master at 3.3V supply voltage. 
    \textbf{The ASIC AMUX output} is buffered by an operational amplifier. Finally,
    \textbf{wire headers} provide connections for signals like I2C, ASIC reset, GND and power.

\subsection{Interface mezzanine boards}
\label{sec:mezzanines}
During the FBCM ASIC testing campaign, several options are required for interfacing to the external measurement and read-out equipment. For this purpose, the test system includes replaceable interface mezzanine cards with different connectors. This also allows new interfacing options in the future without changing the ASIC carrier board. Initially, the mezzanine board types are: \textbf{SMA} break-out of all fast signals, \textbf{miniDP} for interfacing to KSU FMC board~\cite{KSUFMC} (shown in blue in Figure~\ref{fig:test-system-pcb-render}), and 
"electrical link" for interfacing to the adapted CMS Inner Tracker portcard~\cite{Orfanelli:2022zhe} with lpGBT transceiver~\cite{Biereigel:2020jri}.

%% file: tex/testplan.tex
\section{Test plan}
\label{sec:testplan}

The test system is developed to support all testing needs throughout the project lifetime from initial ASIC tests to full read-out chain setup.

\textbf{The ASIC validation and qualification} is initially done separately from the other parts of the system. This involves the measurement of the current consumption and monitoring of the binary and analog signals using an oscilloscope and other measurement equipment via the SMA mezzanine board. The input signal for the FBCM ASIC is emulated using a calibration strobe from a lab signal generator, while capacitors at the ASIC inputs are used for noise measurement. The system is powered from a lab power supply, and the I2C control of the FBCM ASIC is implemented using a PC and a development board.

For the \textbf{initial read-out test}, it is planned to use a galvanic interface with a miniDP mezzanine board for connecting the FBCM ASIC outputs to an FC7 FPGA-based back-end board~\cite{Pesaresi_2015}. This avoids extra complexity in the read-out chain and validates the reception and processing of FBCM output signals in the back-end part of the system. The input signal for the FBCM ASIC will be generated in the laboratory by charged particles from a radioactive source passing through the silicon pad sensors. There is also an option to inject a known charge directly to the ASIC input, called a calibration strobe. 

Finally, \textbf{the full read-out chain test} targets the validation of the full system used in the final application, where all other parts of the system are also added. Most importantly, lpGBT and VTRx+ transceivers are used to send the signals via an optical link from the FBCM to the back-end FPGA-based board. The firmware will decode the data and analyse ToA/ToT parameters to build respective histograms. In addition, powering via DC-DC bPOL12V module~\cite{Faccio:2020rae} will be added, as well as the I2C control of the FBCM ASIC and the calibration strobe generation via the lpGBT downstream channels. This configuration will be used in the planned validation of the system with a test beam.

%% file: tex/summary.tex
\section{Summary}
\label{sec:summary}
The Fast Beam Condition Monitor is a dedicated luminosity and beam-induced background monitor based on silicon-pad sensors with a fast front-end designed for the Phase-2 upgrade of the CMS detector. The FBCM23, a dedicated CMOS 65\,nm radiation tolerant read-out ASIC connected to six silicon-pad sensors is installed in the front-end module of the FBCM system. The ASIC discriminates the signals from the sensors and sends resulting high-speed binary signals via a gigabit optical link to an FPGA-based back-end processing board. The specialized firmware inside the FPGA analyses the signal timing properties (time-of-arrival and time-over-threshold) and builds respective histograms, which enable bunch-by-bunch precision luminosity measurements in real time.

For testing purposes, a specialized system was designed and built to cover validation, qualification and testing needs of the FBCM system throughout all project stages, starting from initial ASIC tests to final full read-out chain system-level tests.

%% file: main.bbl
\providecommand{\href}[2]{#2}\begingroup\raggedright\endgroup